\documentclass[prd,superscriptaddress,onecolumn,showpacs,amsmath,amssymb]{revtex4}
\def\comment#1{}
\usepackage{graphicx}
\usepackage{dcolumn}
\usepackage{bm}
\usepackage{mathrsfs}
\begin{document}
\title{Constructing EOB dynamics with numerical energy flux for intermediate-mass-ratio inspirals}
\author{Wen-Biao Han}
\email{wenbiao@icra.it} \affiliation{ICRANet and Physics Department,
University of Rome ``La Sapienza'', Piazzale della Repubblica 10,
65122, Pescara, Italy}
\author{Zhoujian Cao}
\email{zjcao@amt.ac.cn} \affiliation{Institute of Applied
Mathematics, Academy of Mathematics and Systems Science, Chinese
Academy of Sciences, Beijing 100190, China}
\date{\today}
\begin{abstract}
A new scheme for computing dynamical evolutions and gravitational
radiations for intermediate-mass-ratio inspirals (IMRIs) based on an
effective one-body (EOB) dynamics plus Teukolsky perturbation theory
is built in this paper. In the EOB framework, the dynamics
essentially affects the resulted gravitational waveform for binary
compact star system. This dynamics includes two parts. One is the
conservative part which comes from effective one-body reduction. The
other part is the gravitational back reaction which contributes to
the shrinking process of the inspiral of binary compact star system.
Previous works used analytical waveform to construct this back
reaction term. Since the analytical form is based on post-Newtonian
expansion, the consistency of this term is always checked by
numerical energy flux. Here we directly use numerical energy flux by
solving the Teukolsky equation via the frequency-domain method to
construct this back reaction term.  And the conservative correction
to the leading order terms in mass-ratio is included in the
deformed-Kerr metric and the EOB Hamiltonian. We try to use this
method to simulate not only quasi-circular adiabatic inspiral but
also the nonadiabatic plunge phase. For several different spinning
black holes, we demonstrate and compare the resulted dynamical
evolutions and gravitational waveforms.

\end{abstract}

\maketitle
\section{Introduction}
Intermediate-mass-ratio inspirals (IMRIs): compact objects with
stellar mass spiralling into intermediate mass ($50 \leq M \leq
1000M_\odot$) black holes are expected to be a key source for LIGO,
VIRGO and the future Einstein telescope. It was estimated that
Advanced LIGO can see 3-30 such events per year out to distances of
several hundred Mpc\cite{Brown}. Due to a small mass ratio of IMRIs
($\mu/M \sim 10^{-2}-10^{-3}$), the small body lingers in the large
black hole¡¯s strong-curvature region for about hundreds wave cycles
before merger, and then produce a good signal-to-noise ratio.
Considering the waveform-match technology of LIGO's data analysis,
ones should first model enough accurate and numerous waveform
templates for IMRI sources.

The post-Newtonian approximation has completely broken down in this
high relativistic regime. Up to now, numerical relativity still
encounters some difficulties to simulate the binary black hole
systems when the mass ratio becomes small. The newest result
achieves the mass ratio $0.01$ (without spin) only with two orbit's
evolution\cite{NR1}.

For extreme-mass-ratio inspirals (EMRIs), the mass ratio about
$10^{-4}-10^{-7}$, the small objects can be treated as a small
perturbation to the large black hole¡¯s gravitational field and can
be modeled by black-hole perturbation theory built by Regge, Wheeler
and Zerilli in Schwarzschild spacetime\cite{R-W,Zerilli} and by
Teukolsky in Kerr background\cite{Teukolsky1,Teukolsky2}. For
example, a successful approach is that the orbit of small body is
just geodesics of test particle in the background of central black
hole, and the spiral of small object is simulated as a sequence of
adiabatically shrinking geodesics by changing the constants of
motion due to gravitational
radiation\cite{Poisson1,Poisson2,Shibata,Hughes1,Hughes2,Glampedakis1,Hughes3,Hughes4,Hinderer,Fujita3}.
But, when mass ratio increases, for IMRIs, the small object cannot
be treated as test particle and the self-gravitational potential of
small body becomes considerable. And perhaps the linear perturbation
approximation is not valid in this case.

In last decade, using effective one-body (EOB) resummed analytical
formalism
\cite{Buonanno1,Buonanno2,Damour1,Damour2,Buonanno3,Damour3,Pan1} to
develop accurate analytical models of dynamics and waveforms of
black hole binaries has gotten some exciting
results\cite{Buonanno4,Damour4,Damour5,Damour6,Boyle,Buonanno5,Damour7,Pan2,Barausse1}.
Recently, some researches used EOB to model the dynamical evolution
of IMRIs and extract waveforms via a multipolar
Regger-Wheeler-Zerilli type perturbation
approach\cite{Nagar,Bernuzzi1,Bernuzzi2}. And an effort of
constructing the EOB waveform models for EMRIs by calibration of
several high-order post-Newtonian parameters was also presented
\cite{Yunes1,Yunes2}.

The framework of effective one-body has two building blocks which
are dynamics and the corresponding waveform. But the dynamics
essentially affects the resulted gravitational waveform for binary
compact star system. This dynamics includes two parts. One is the
conservative part which comes from effective one-body reduction. The
other part is the gravitational back reaction which contributes to
the shrinking process of the inspiral of binary compact star system.
Previous works used analytical waveform to construct this back
reaction term. Since the analytical form is based on some
assumptions, the consistency of this term is always checked by
numerical energy flux. Differently we directly use numerical energy
flux to construct this back reaction term in this work.

Considering that the perturbation method works well for intermediate
mass ratio binaries\cite{Lousto1,Lousto2}, in this paper, we model
the gravitational wave sources of IMRIs by combining the EOB frame
with the Teukolsky perturbation-based energy fluxes. Thanks to the
EOB theory, we treat the binary system by an effective particle mass
$\mu=m_1 m_2/(m_1+m_2)$ and a deformed Kerr (or Schwarzschild if
without spin) spacetime with mass $M=m_1+m_2$ where $m_1$ is the
mass of massive black holes and $m_2$ the mass of small body. The
gravitational energy fluxes and waveforms are calculated by solving
the Teukolsky equation in frequency domain not by the EOB fluxes and
waveforms, as well as the inspiral dynamics is computed not
following the adiabatic approximation but by the EOB dynamics which
was developed in Refs.\cite{Nagar,Bernuzzi1,Bernuzzi2}.

We must emphasize here that the scheme we used in this paper is
different with the works by Yunes et al. which used the Teukolsky
fluxes to calibrate EOB waveforms\cite{Yunes1,Yunes2}, and also
different with the references \cite{Nagar,Bernuzzi1,Bernuzzi2} which
use perturbation method to read the waveforms but not to radiation
reaction. Our calculation covers both quasi-circular inspiral and
nonadiabatic plunge without merger and ringdown stages. The results
in this paper should be important for checking the EOB PN-expanded
energy flux. With our method, one can obtain information for IMRIs
which is not yet been accessible by full numerical relativity.
Moreover, our method may be expanded to elliptic and even
non-equatorial orbits. In the future works of this series, we will
expand this calculation to solve Teukolsky equation in time domain
and model the entire process of IMRIs including merger and ringdown.

This paper is organized as follows. In the next section, we
introduce EOB Hamiltonian and dynamics briefly. Then we describe our
numerical technologies used in solving the Teukolsky equation to
 obtain fluxes and waveforms. In Sec. IV we present our calculation
 results for several IMRI cases. Finally, conclusion and discussion
 of this work are given in the last section.

We use units $G=c=1$ and the metric signature $(-,+,+,+)$. Distance
and time are measured by the effective Kerr black-hole mass $M$.

\section{EOB Hamilton with deformed Kerr background}
The well-known EOB formalism was first introduced by Buonanno and
Damour about ten years ago to model comparable-mass black hole
binaries\cite{Buonanno1,Buonanno2}, and was also applied in small
mass-ratio systems\cite{Nagar,Bernuzzi1,Bernuzzi2,Damour8}. Here we
assume an IMRI system with central Kerr black hole $m_1$ and
inspiralling object $m_2$ which is restricted on the equatorial
plane of $m_1$, total mass $M=m_1+m_2$, reduced mass $\mu=m_1m_2/M$
and symmetric mass ratio $\nu=\mu/M$. Here we don't consider the
spin of the small object. Then the deformed Kerr metric takes the
form \cite{Barausse1}
\begin{align}
g^{tt}&=-\frac{\Lambda_t}{\Delta_t \Sigma},\\
g^{rr}&=\frac{\Delta_r}{\Sigma},\\
g^{\theta\theta}&=\frac{1}{\Sigma},\\
g^{\phi\phi}&=\frac{1}{\Lambda_t}\left(-\frac{\widetilde{\omega}_{fd}^2}{\Delta_t\Sigma}+\frac{\Sigma}{\sin^2\theta}\right),\\
g^{t\phi}&=-\frac{\widetilde{\omega}_{fd}}{\Delta_t\Sigma},
\end{align}
where
\begin{align}
\Sigma &=r^2+a^2\cos^2{\theta},\\
\Delta_t&=r^2[A(u)+\frac{a^2}{M^2}u^2], \\
\Delta_r&=\Delta_t D^{-1}(u),\\
\Lambda_t&=(r^2+M^2a^2)^2-M^2a^2\Delta_t,\\
\widetilde{\omega}_{fd}&=2M a r+\omega^{fd}_1
\nu\frac{aM^3}{r}+\omega^{fd}_2\nu\frac{Ma^3}{r},
\end{align}
where $u=M/r$ and
\begin{align}
A(u)&=1-2u+2\nu u^3+\nu(\frac{94}{3}-\frac{41}{32}\pi^2)u^4, \\
D^{-1}(u)&=1 + \log[1 + 6\nu u^2 + 2(26-3\nu)\nu u^3].
\end{align}
Note that $a$ is the so-called deformed-Kerr spin parameter, which
is not exactly equal to the spin parameter of Kerr black hole
itself. The values of $\omega^{fd}_1$ and $\omega^{fd}_2$ given by a
preliminary comparison of EOB model with numerical relativity
results are about -10 and 20 respectively\cite{Barausse2,Rezzolla}.

The EOB Hamiltonian takes the form
\begin{align}
H_{\text{EOB}}=M\sqrt{1+2\nu(H_{\text{eff}}/\mu-1)}.
\end{align}
The effective Hamiltonian $H_{\text{eff}}$ in this paper is just the
Hamiltonian of a nonspinning (NS) particle in the deformed-Kerr
metric
\begin{align}
H_{\text{NS}}=\beta^i p_i+\alpha\sqrt{\mu^2+\gamma^{ij}p_i
p_j},\label{HNS}
\end{align}
and
\begin{align}
\alpha&=\frac{1}{\sqrt{-g^{tt}}},\\
\beta^i &=\frac{g^{ti}}{g^{tt}},\\
\gamma^{ij}&=g^{ij}-\frac{g^{ti}g^{tj}}{g^{tt}}.
\end{align}
With all of this, the effective Hamiltonian(\ref{HNS}) can be
written as
\begin{align}
H_{\text{NS}}=p_\phi\frac{\widetilde{\omega}_{fd}}{\Lambda_t}+\left(\frac{\Delta_t
\Sigma}{\Lambda_t}\right)^{1/2}\sqrt{\mu^2+\frac{\Sigma}{\Lambda_t\sin^2{\theta}}p_\phi^2+\frac{\Delta_r}{\Sigma}p_r^2}.
\end{align}

The EOB dynamical evolution equations under radiation reaction can
be given as
\begin{align}
\dot{r}&=\frac{\Delta_t \Delta_r p_r}{H_{\text{EOB}}H_r\Lambda_t},\label{rdot}\\
\dot{\phi}&=\frac{\Delta_t \Sigma^2}{\Lambda_t^2}\frac{p_\phi}{H_{\text{EOB}}H_r}\equiv \Omega_\phi,\label{fdot}\\
\dot{p_r}&=(\partial H_{\text{NS}}/\partial r)/H_{\text{EOB}},\label{prdot}\\
\dot{p_\phi}&=\hat{\mathcal{F}_\phi},\label{pfdot}
\end{align}
where
$H_r=H_{\text{NS}}-p_\phi\frac{\widetilde{\omega}_{fd}}{\Lambda_t}$.
$\hat{\mathcal{F}_\phi}$ (of order $O(\nu)$) represents the non
conservative radiation reaction force. Following the expression
given in \cite{Bernuzzi1,Bernuzzi2,Damour9}, we have
\begin{align}
\hat{\mathcal{F}_\phi}\equiv -\frac{32}{5}\nu\Omega^5 r^4
\hat{f}_{\text{DIN}}\label{rrforce},
\end{align}
where $\hat{f}_{\text{DIN}}=F^{l_{\text{max}}}/F_\text{Newt}$ is the
Newton normalized energy flux up to multipolar order $l_\text{max}$.
In the EOB framework, $F^{l_{\text{max}}}$ is given by ``improved
resummation'' technique of Ref.\cite{Damour9} for a nonspinning
binary and Ref.\cite{Pan1} for a spinning one. In this paper,
 we use energy flux calculated by the Teukolsky perturbation
 theory instead of EOB energy flux in Eq.(\ref{pfdot}). As discussed in Ref.\cite{Bernuzzi1}, the EOB dynamical
 equations (\ref{rdot}-\ref{pfdot}) do not make any adiabatic
 approximation which is extensively used to model EMRIs by solving
 Teukolsky equation \cite{Hughes1,Hughes2,Glampedakis1,Hughes3,Hughes4}.

\section{Teukolsky equation and numerical calculation}
\subsection{The Teukolsky equation}
The Teukolsky formalism considers perturbation on the Weyl curvature
scale $\psi_4$ instead of metric perturbation like the
Regge-Wheeler-Zerilli equations. For Schwarzschild black holes these
two formalisms are equivalent, but Teukolsky formalism facilitates
us to study super-massive spinning black hole. Different to time
domain method in \cite{Nagar,Bernuzzi1,Bernuzzi2}, we decompose
$\psi_4$ in the frequency domain~\cite{Teukolsky1},
\begin{align}
\psi_4=\rho^4\int^{+\infty}_{-\infty}{d\omega\sum_{lm}{R_{lm\omega}(r)_{~-2}S^{a\omega}_{lm}(\theta)e^{im\phi}e^{-i\omega
t}}},
\end{align}
where $\rho=-1/(r-ia\cos{\theta})$. The function $R_{lm\omega}(r)$
satisfies the radial Teukolsky equation
\begin{align}
\Delta^2\frac{d}{dr}\left(\frac{1}{\Delta}\frac{d
R_{lm\omega}}{dr}\right)-V(r)R_{lm\omega}=-\mathscr{T}_{lm\omega}(r),\label{Teukolsky}
\end{align}
where $\mathscr{T}_{lm\omega}(r)$ is the source term, and the
potential is
\begin{align}
V(r)=-\frac{K^2+4i(r-M)K}{\Delta}+8i\omega r+\lambda,
\end{align}
where $K=(r^2+a^2)\omega-ma, ~\lambda=E_{lm}+a^2\omega^2-2a m w-2$.
The spin-weighted angular function $_{-2}S^{a\omega}_{lm}(\theta)$
obeys the following equation,
\begin{align}
\frac{1}{\sin{\theta}}\frac{d}{d\theta}\left(\sin{\theta
\frac{d_{~-2}S^{a\omega}_{lm}}{d\theta}}\right)+\left[(a\omega)^2\cos^2{\theta}+4a\omega\cos{\theta}-\left(\frac{m^2-4m\cos\theta+4}{\sin^2\theta}\right)+E_{lm}\right]
~_{-2}S^{a\omega}_{lm}=0.\label{spinweightedspheroidal}
\end{align}
The radial Teukolsky Eq.(\ref{Teukolsky}) has the general solution
\begin{align}
R_{lm\omega}(r)=\frac{R^{\infty}_{lm\omega}(r)}{2i\omega
B^{\text{in}}_{lm\omega}D^{\infty}_{lm\omega}}\int^{r}_{r_+}{dr'\frac{R^\text{H}_{lm\omega}(r')\mathscr{T}_{lm\omega}(r')}{\Delta(r')^2}}+\frac{R^{\text{H}}_{lm\omega}(r)}{2i\omega
B^{\text{in}}_{lm\omega}D^{\infty}_{lm\omega}}\int^{\infty}_{r}{dr'\frac{R^\infty_{lm\omega}(r')\mathscr{T}_{lm\omega}(r')}{\Delta(r')^2}},\label{solutiont}
\end{align}
where the $R^{\infty}_{lm\omega}(r)$ and $R^{H}_{lm\omega}(r)$ are
two independent solutions of the homogeneous Teukolsky equation.
They are chosen to be the purely ingoing wave at the horizon and
purely outgoing wave at infinity, respectively,
\begin{align}\nonumber
R^{\text{H}}_{lm\omega}&=B^{\text{hole}}_{lm\omega}\Delta^2
e^{-ipr*},\quad
r\rightarrow r_+\\
R^{\text{H}}_{lm\omega}&=B^{\text{out}}_{lm\omega}r^3 e^{i\omega
r*}+r^{-1}B^{\text{in}}_{lm\omega}r e^{-i\omega r*},\quad
r\rightarrow \infty;
\end{align}
and
\begin{align}\nonumber
R^{\infty}_{lm\omega}&=D^{\rm{out}}_{lm\omega} e^{ip r*}+\Delta^2
D^{\text{in}}_{lm\omega}r e^{-ip r*},\quad
r\rightarrow r_+\\
R^{\infty}_{lm\omega}&=r^3 D^{\infty}_{lm\omega} e^{-i\omega
r*},\quad r\rightarrow \infty,
\end{align}
where $k=\omega-ma/2Mr_+$ and $r^*$ is the ``tortoise coordinate".
The solution (\ref{solutiont}) must be purely ingoing at horizon and
purely outgoing at infinity. That is,
\begin{align}
R_{lm\omega}(r\rightarrow \infty)&=Z^{\text{H}}_{lm\omega}r^3
e^{i\omega
r*},\\
R_{lm\omega}(r\rightarrow r_+)&=Z^{\infty}_{lm\omega}\Delta^2 e^{-ip
r*}.
\end{align}
The complex amplitudes $Z^{\text{H},\infty}_{lm\omega}$ are defined
as
\begin{align}\nonumber
Z^{\text{H}}_{lm\omega}&=\frac{1}{2i\omega
B^{\text{in}}_{lm\omega}}\int^{r}_{r_+}{dr'\frac{R^\text{H}_{lm\omega}(r')\mathscr{T}_{lm\omega}(r')}{\Delta(r')^2}},\\
Z^{\infty}_{lm\omega}&=\frac{B^{\text{hole}}_{lm\omega}}{2i\omega
B^{\text{in}}_{lm\omega}D^{\infty}_{lm\omega}}\int^{\infty}_{r}{dr'\frac{R^\infty_{lm\omega}(r')\mathscr{T}_{lm\omega}(r')}{\Delta(r')^2}}.\label{Z1}
\end{align}

The particle in this paper is in circular orbit on the equatorial
plane, thus the particle's motion is described only as the harmonic
of the frequency $\Omega_\phi$ (\ref{fdot}). We define
\begin{align}
\omega_m=m\Omega_\phi.\label{harmonicfrequency}
\end{align}
Then $Z^{H,\infty}_{lm\omega}$ are decomposed as
\begin{align}\nonumber
Z^{\text{H}}_{lm\omega}&=Z^{\text{H}}_{lm}\delta(\omega-\omega_m),\\
Z^{\infty}_{lm\omega}&=Z^{\infty}_{lm}\delta(\omega-\omega_m).\label{Z2}
\end{align}
The amplitudes $Z^{H,\infty}_{lm\omega}$ fully determine the energy
fluxes of gravitational radiations,
\begin{align}
\dot{E}^{\infty,\text{H}}&=\sum_{lm\omega}{\frac{|Z^{\mathrm{H},\infty}_{lm\omega}|^2}{4\pi\omega^2_{m}}},\label{energyfluxf}
\end{align}
and the gravitational waveform:
\begin{align}
h_+-ih_\times=\frac{2}{r}\sum_{lm}{\frac{Z^{\mathrm{H}}_{lm\omega}}{\omega^2_{m}}S^{a\omega_{m}}_{lm}(\theta)e^{-i\omega_{m}t+im\phi}}.
\end{align}

In the next subsection, we briefly introduce the traditional
numerical technique solving the Teukolsky equation by
Sasaki-Nakamura translation\cite{Sasaki}. And in Sec. III C, we
introduce a new numerical method solving the Teukolsky equation by
Fujita and Tagoshi\cite{Fujita1,Fujita2}.

\subsection{The Sasaki-Nakamura equation}
As mentioned in the above subsections, in order to get
$Z^{H,\infty}_{lm}$, we should integrate the homogenous version of
Eq.(\ref{Teukolsky}). But there is a difficulty when one numerically
integrates Eq.(\ref{Teukolsky}) due to the long-range nature of the
potential $V(r)$ in (\ref{Teukolsky}). In order to solve this
problem, Sasaki and Nakamura developed the Sasaki-Nakamura function
$X(r)$, governed by a short-ranged potential, to replace the
Teukolsky function $R(r)$ \cite{Sasaki}. The Sasaki-Nakamura
equation reads as
\begin{align}
\frac{d^2
X_{lm\omega}}{dr*^2}-F(r)\frac{dX_{lm\omega}}{dr*}-U(r)X_{lm\omega}=0.\label{s-keq}
\end{align}
The functions $F(r),~U(r)$ can be found in Ref.\cite{Sasaki}. The
Sasaki-Nakamura equation also admits two asymptotic solutions,
\begin{align}
X^{H}_{lm\omega}&=e^{-ipr*},\quad r\rightarrow r_+,\label{snhh}\\
X^{H}_{lm\omega}&=A^{\rm{out}}_{lm\omega}\bar{P}(r)e^{i\omega
r*}+A^{\rm{in}}_{lm\omega}P(r)e^{-i\omega r*},\quad r\rightarrow
\infty;\label{snh8}
\end{align}
and
\begin{align}
X^{\infty}_{lm\omega}&=C^{\rm{out}}_{lm\omega}\bar{P}(r)e^{ip
r*}+C^{\rm{in}}_{lm\omega}P(r)e^{-ip r*},\quad r\rightarrow r_+,\label{sn8h}\\
X^{\infty}_{lm\omega}&=\bar{P}(r)e^{-i\omega r*},\quad r\rightarrow
\infty,\label{sn88}
\end{align}
where $P(r),~\bar{P}(r)$ can be found in Refs.\cite{Dolan,Hughes5}.

The solution of Eq.(\ref{s-keq}) is transformed to the solution of
the Teukolsky equation by
\begin{align}
R^{H,\infty}_{lm\omega}=\frac{1}{\eta}\left[\left(\alpha+\frac{\beta_{,r}}{\Delta}\right)\frac{\Delta
X^{H,\infty}_{lm\omega}}{\sqrt{r^2+a^2}}-\frac{\beta}{\Delta}\frac{d}{dr}\frac{\Delta
X^{H,\infty}_{lm\omega}}{\sqrt{r^2+a^2}}\right].
\end{align}
The relations between the coefficients of the Sasaki-Nakamura
function and the Teukolsky function are
\begin{align}
B^{\rm{in}}_{lm\omega}=-\frac{A^{\rm{in}}_{lm\omega}}{4\omega^2},~B^{\rm{hole}}_{lm\omega}=\frac{1}{d_{lm\omega}},~D^{\infty}_{lm\omega}=-\frac{4\omega^2}{c_0},
\end{align}
and the functions $\alpha,~\beta, ~\eta,$ and $d_{lm\omega}$ are
given explicitly in Ref.\cite{Sasaki} and in Appendix B of
\cite{Hughes1}.

Our numerical scheme to integrate the Sasaki-Nakumura equation has
been used in Ref.\cite{Han1} for spinning test particle. The
integrator we adopted is Runge-Kutta 7(8) and a variant of
Richardson extrapolation is implemented to accurately compute the
value of $X_{lm}^{H,\infty}$ in Eqs.(\ref{snhh}-\ref{sn88})
approaching to infinity and horizon. The calculation of
spin-weighted spheroidal harmonics in
Eq.(\ref{spinweightedspheroidal}) and more details can be founded in
Ref.\cite{Hughes1}.

\subsection{The semi-analytical method for Teukolsky equation}
Our main numerical technique in this work is a semi-analytical
numerical method developed by Fujita and
Tagoshi\cite{Fujita1,Fujita2} which is based on the MST (Mano,
Suzuki and Takasugi) analytical solutions of homogeneous Teukolsky
equation\cite{MST,Sasaki2}. In the MST method, the homogeneous
solutions of Teukolsky equation are expressed in terms of two kinds
of series of special functions. The first one consists of series of
hypergeometric functions and is convergent at the horizon
\begin{align}
R_{lm\omega}^{\text{in}}=e^{i\epsilon\kappa
x}(-x)^{-s-i(\epsilon+\tau)/2}(1-x)^{i(\epsilon-\tau)/2}p_{\text{in}}(x),
\end{align}
where $p_{\text{in}}$ is expanded in a series of hypergeometric
functions as
\begin{align}
p_{in}(x)=\sum_{n=-\infty}^{\infty}{a_n F(n+\nu+1-i\tau,
-n-\nu-i\tau;1-s-i\epsilon-i\tau;x)},\label{hypergeo}
\end{align}
and $x=\omega(r_+-r)/\epsilon\kappa$,
$\epsilon=2M\omega$,$\kappa=\sqrt{1-a^2}$,$\tau=(\epsilon-m
a)/\kappa$. The hypergeometric function $F(\alpha,\beta;\gamma;x)$
can be found in mathematic handbooks.

The second one consists of series of Coulomb wave functions which is
convergent at infinity. The homogeneous solution of Teukolsky
equation is
\begin{align}
R_C=z^{-1-s}(1-\epsilon\kappa/z)^{-s-i(\epsilon+\tau)/2}f_\nu(z),
\end{align}
where $f_\nu(z)$ is expressed in a series of Coulamb wave functions
as
\begin{align}
f_\nu(z)=\sum_{-\infty}^{\infty}{(-i)^n\frac{(\nu+1+s-i\epsilon)_n}{(\nu+1-s+i\epsilon)_n}a_nF_{n+\nu}(-is-\epsilon,z)},\label{coulomb}
\end{align}
and $z=\omega(r-r_-),(a)_n=\Gamma(a+n)/\Gamma(a)$, $F_N(\eta,z)$ is
a Coulomb wave function. Notice that both
Eqs.(\ref{hypergeo},\ref{coulomb}) involve a parameter $\nu$, the
so-called renormalized angular momentum.

The key part of Fujita and Tagoshi's method is searching the
renormalized angular momentum numerically. When harmonic frequency
$\omega$ is under some critical values, real solutions of $\nu$ can
be found\cite{Fujita1}. If $\omega$ exceeding these critical values,
complex solutions of $\nu$ can also be found\cite{Fujita2}. But when
$\omega$ becomes very large, it becomes very difficult to search the
solution of $\nu$ accurately. In this paper, we mainly employ the
new method by Fujita and Tagoshi to solve the Teukolsky equation
because of its high accuracy and high efficiency, but when $\omega$
becomes large and difficult to find $\nu$ accurately, we use the
numerical integration method to evaluate the Teukolsky equation.

For checking our numerical code's validity, in Table \ref{table1}
and Table \ref{table2}, we list our results by two kind of methods
and the data published by Ref.\cite{Fujita1}. We find our results
are almost same with Fujita and Tagoshi's, and the accuracy of our
numerical integration of the Teukolsky equation is also good.

\begin{table}
\caption{The gravitational wave luminosity at infinity for $r=6M$
and $a=0.9$ (The data labeled ``Ref.\cite{Fujita1}'' are kindly
offered by Dr. Fujita).}
\label{table1}
\begin{tabular}{c c|c c| c c }
\hline \hline $l$ & $|m|$ & Ref.\cite{Fujita1} & This paper & Numerical integration & Relative error \\
\hline 2&2& $4.61839129214686\times10^{-4}$&$4.61839129214674\times
10^{-4}$&$4.6183912944\times
10^{-4}$&$4.98\times 10^{-10}$\\
2&1& $6.69474358662814\times 10^{-7}$ &$6.69474358662808\times
10^{-7}$&$6.6947435969\times 10^{-7}$&$1.49\times 10^{-9}$\\
3&3& $8.03430093731846\times 10^{-5}$ &$8.03430093731833\times
10^{-5}$&$8.0343009232\times 10^{-5}$&$1.76\times 10^{-9}$\\
3&2& $2.92125227685241\times 10^{-7}$ &$2.92125227685261\times
10^{-7}$&$2.9212522761\times 10^{-7}$&$2.40\times 10^{-10}$\\
3&1& $4.30107466504698\times 10^{-9}$ &$4.30107466504728\times
10^{-9}$&$4.3010746567\times 10^{-9}$&$1.86\times 10^{-9}$\\
4&4& $1.72937852645569\times 10^{-5}$ &$1.72937852645567\times
10^{-5}$&$1.7293785258\times 10^{-5}$&$4.05\times 10^{-10}$\\
5&5& $4.01481595546757\times 10^{-6}$ &$4.01481595546768\times
10^{-6}$&$4.0148159516\times 10^{-6}$&$9.71\times 10^{-10}$\\
6&6& $9.65798420846838\times 10^{-7}$ &$9.65798420846792\times
10^{-7}$&$9.6579842187\times 10^{-7}$&$1.96\times 10^{-9}$\\
\hline \hline
\end{tabular}
\end{table}

\begin{table}
\caption{The gravitational wave luminosity at horizon for $r=6M$ and
$a=0.9$ (The data labeled ``Ref.\cite{Fujita1}'' are kindly offered
by Dr. Fujita).} \label{table2}
\begin{tabular}{c c|c c| c c }
\hline \hline $l$ & $|m|$ & Ref.\cite{Fujita1} & This paper & Numerical integration & Relative error \\
\hline 2&2&
$-3.98206695526010\times10^{-6}$&$-3.98206695525998\times
10^{-6}$&$-3.9820668046\times
10^{-6}$&$3.77\times 10^{-8}$\\
2&1& $-7.10533640284950\times 10^{-8}$ &$-7.10533640284937\times
10^{-8}$&$-7.1053351261\times 10^{-8}$&$1.79\times 10^{-7}$\\
3&3& $-1.17015834099274\times 10^{-7}$ &$-1.17015834099273\times
10^{-7}$&$-1.1701585797\times 10^{-7}$&$2.05\times 10^{-7}$\\
3&2& $-2.54494163085054\times 10^{-9}$ &$-2.54494163085075\times
10^{-9}$&$-2.5449398939\times 10^{-9}$&$6.68\times 10^{-7}$\\
3&1& $-1.41731814300288\times 10^{-10}$ &$-1.41731814300298\times
10^{-10}$&$-1.4173182794\times 10^{-10}$&$9.86\times 10^{-8}$\\
4&4& $-4.26087026755638\times 10^{-9}$ &$-4.26087026755628\times
10^{-9}$&$-4.2608720408\times 10^{-9}$&$3.99\times 10^{-7}$\\
5&5& $-1.66087471582563\times 10^{-10}$ &$-1.66087471582562\times
10^{-10}$&$-1.6608748796\times 10^{-10}$&$9.64\times 10^{-8}$\\
6&6& $-6.66108480342866\times 10^{-12}$ &$-6.66108480342902\times
10^{-12}$&$-6.6610834965\times 10^{-12}$&$4.80\times 10^{-7}$\\
\hline \hline
\end{tabular}
\end{table}

\section{Relativistic dynamics and waveforms}
The mass ratio of IMRI systems we adopted is $0.01$, and the central
massive body is a Kerr black hole . The initial data ($p_r,
~p_\phi$) for solving Eqs.(\ref{rdot}-\ref{pfdot}) follows the
so-called post-circular initial conditions described in
\cite{Damour1} and take limit $\nu\rightarrow 0$,
\begin{align}
p_\phi&=L_z,\\
p_r&=-[\frac{\hat{\mathcal{F}}}{dL_z/dr}]\dot{t}\Sigma/\Delta.
\end{align}
We should point out here that usually we do not take the limit
$\nu\rightarrow 0$ (which was used in
Refs.\cite{Nagar,Bernuzzi1,Bernuzzi2}) during the dynamical
evolution; instead, we use the full EOB dynamics described in
Eqs.(\ref{rdot}-\ref{pfdot}). This makes it so we can contain the
first order conservative self-force of the small body in dynamics.
We list our numerical algorithm here for clarity:

1. We calculate the quasi-circular data for the small body at
time-step $t_i$;

2. Use the orbital frequency from the step 1 to solve the Teukolsky
equation in frequency domain by the numerical methods we introduced
in Sec.III up to $l_{\text{max}}=12$;

3. Then get the flux and waveforms (Teukolsky-based) by summing over
the multipoles at the time-step $t_i$;

4. Calculate the dynamics of the small body for the time-step
$t_{i+1}$ by solving the EOB dynamics equations
(\ref{rdot}-\ref{pfdot});

5. Iterate from the step 1.

The method of modeling back reaction and waveforms based on the
Teukolsky equation in frequency domain are quite successful in
quasi-periodic inspirals of extreme mass-ratio cases. It is a good
standard for checking other methods. For example, in
Ref.\cite{Yunes2}, Yunes et al. used the Teukolsky waveforms to
construct more accurate EOB waveform models. But while the mass
ratio $\mu/M\sim 10^{-2}$, the precision of Teukolsky energy fluxes
and waveforms would decrease because it is first order perturbation
theory and also the subleading terms in the mass-ratio introduce
conservative correction is not included in the Teukolsky equation.
It is similar in Refs.\cite{Nagar,Bernuzzi1,Bernuzzi2} where they
calculated $\hat{f}_{\text{DIN}}$ by the EOB analytical radiation
reaction in the $\nu\rightarrow0$ limit and gravitational waves by
the Regge-Wheeler-Zerilli perturbation equation. Anyway, there are
several reasons drive us choose the Teukolsky-based flux to replace
the EOB analytical one to calculate the radiation reaction in this
work. One is that our Teukolsky-based flux does not include any low
velocity approximation comparing with the post-Newtonian expanded
EOB flux. But we also notice that the $\hat{f}_{\text{DIN}}$ based
PN resummed waveform extends effectively the PN waveform beyond the
slow velocity approximation quite well \cite{Fujita4}. One is that
the Teukolsky-based flux can be calculated to arbitrary multipole
(but actually because of a computation limit, we compute to
$l_\text{max}=12$) against the EOB's maximum $l=8$ until now. The
last one is that the black hole absorbed term is naturally included
in the Teukolsky-based energy flux.

There is also a problem to evolve the plunge orbit inside the Kerr
or Schwarzschild ISCO(innermost stable circular orbit) with the
adiabatic approximation frequency-domain code. The plunge orbit is
unstable and do not have a well-defined frequency spectrum. But as
same as the EOB-based evolution of plunge orbit, we can just use the
EOB dynamical equation (\ref{fdot}) to define the orbit frequency
and calculate the harmonic frequency in
Eq.(\ref{harmonicfrequency}). We think this is an alternative way to
calculate the energy flux in the Teukolsky frequency-domain frame.
And the reliability of this definition of orbital frequency was
investigated in \cite{Bernuzzi2}. The key point is that the orbital
evolution in this paper is based on the EOB dynamics
(\ref{rdot}-\ref{pfdot}) but not the adiabatic approximation.

\begin{figure}
\begin{center}
\includegraphics[height=2.5in]{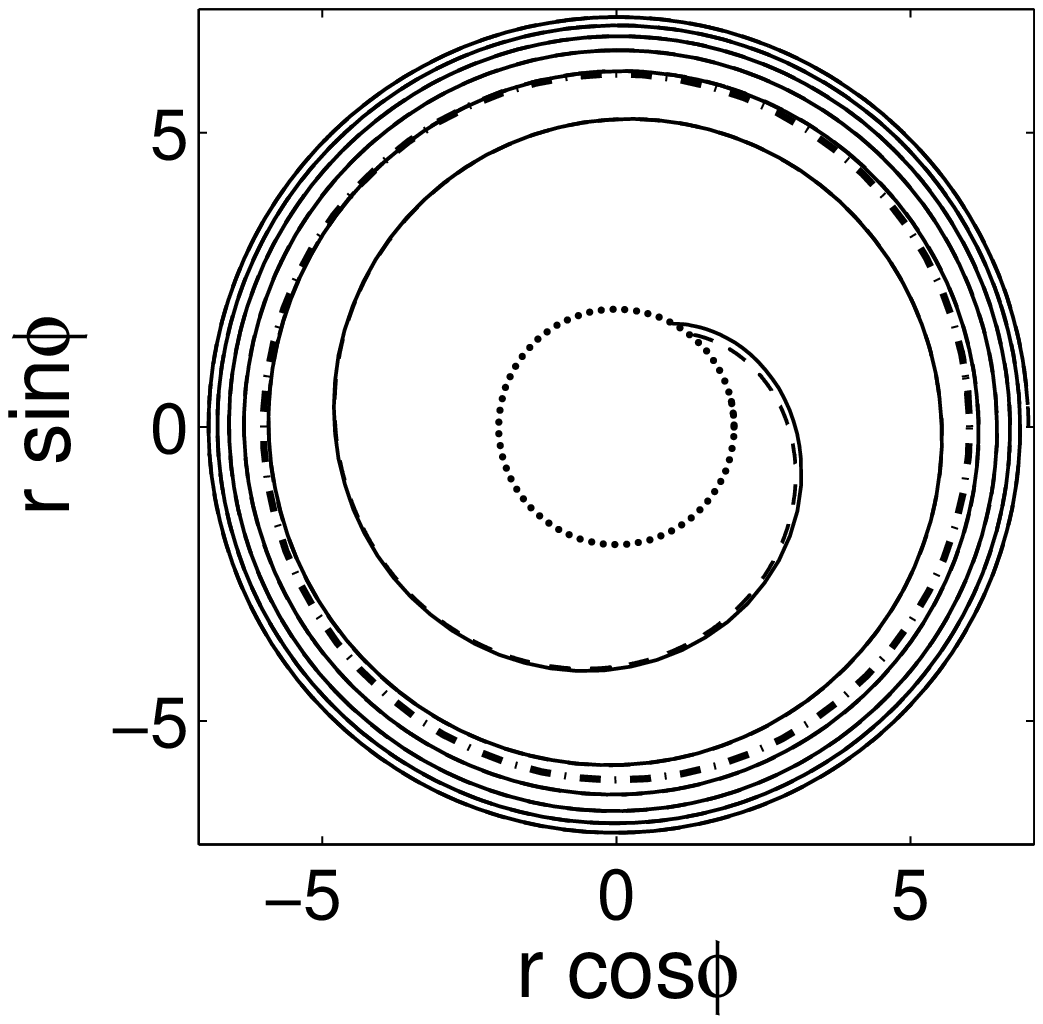}
\includegraphics[height=2.5in]{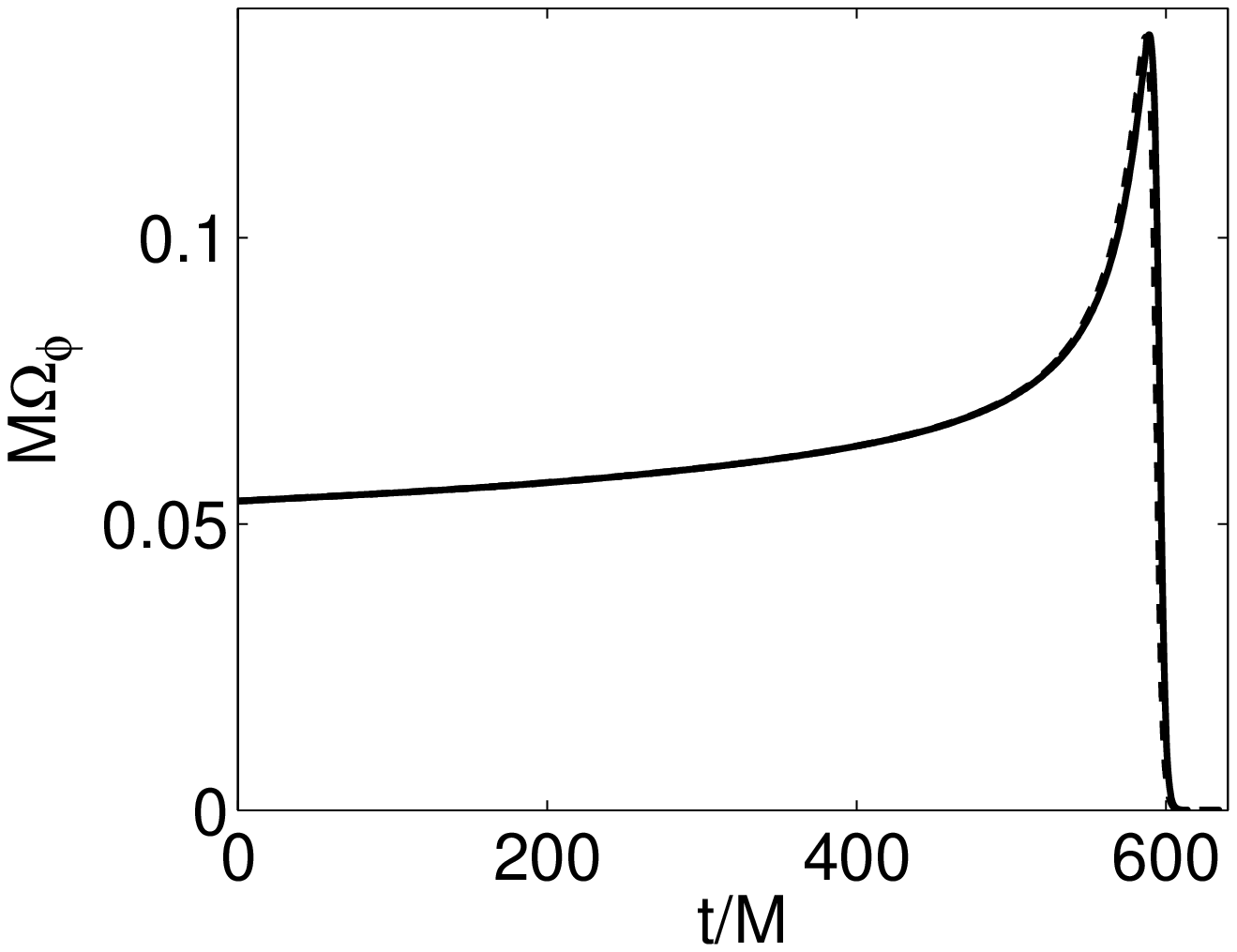}
\caption{Comparison of our method(solid line) with EOB (dashed line,
the data are calculated by Dr. Bernuzzi). Left panel: Transition
from quasi-circular inspiral orbit to plunge from $r_0=7M$ and
$\nu=0.01$. The dashed-dotted line circle is the ISCO at
$r_\text{ISCO}=6M$ and the dotted circle the horizon of black hole;
Right panel: Orbital frequency $\Omega_\phi$ versus coordinate
time.} \label{fig_Schwz}
\end{center}
\end{figure}

First, we compare the same situation in Fig.1 of Ref.\cite{Nagar} by
our numerical algorithm with the 5PN resummed $\hat{f}_{\text{DIN}}$
in \cite{Bernuzzi1,Bernuzzi2}. For comparison, we take the
$\nu\rightarrow 0$ limit as same as the Eqs.(2-6) of
\cite{Bernuzzi1}. Our results are showed in Figure \ref{fig_Schwz},
in which both results are consistently very good at qualitative
level. Because of the different methods for getting energy fluxes,
our results of orbital dynamics and waveforms are slightly
different, with the one by the 5PN resummed back reaction at
quantitative level. In details, before ISCO, during the inspiral,
both results are almost the same, but in plunge phase, a small but
visible dephasing (about 0.16 radian) appears. Of course, this
dephasing should be larger if involving more orbit evolution.

\comment{ The possible reason for this good matching and dephasing
is that Ref.\cite{Nagar} used a 2.5 PN ``factored flux function''
$\hat{f}_\text{DIS}$ of \cite{Damour10} which has a good convergence
toward the exact result known numerically in the $\nu\rightarrow 0$
limit during the quasi-circular inspiral phase but was not checked
in the plunge process. We can see in the right panel of
Figure\ref{fig_Schwz}, before ISCO, the orbital frequency increases
very slow. But after ISCO, the frequency increases very fast and
then the orbital velocity becomes quite large, this makes the 2.5 PN
waveform not accurate enough. }

\begin{figure}
\begin{center}
\includegraphics[width=3.0in]{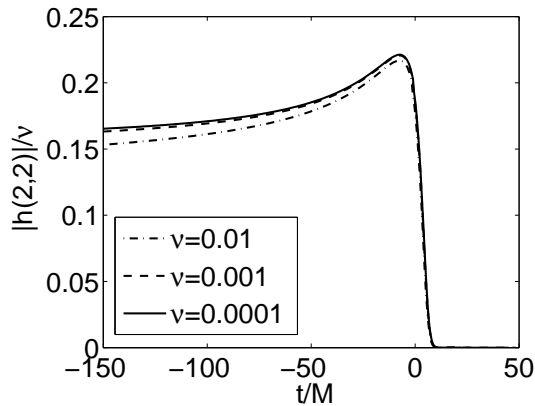}
\caption{``Convergence'' of the waveform when $\nu\rightarrow 0$ for
$a=0$. The values of $\nu$ are $0.01,~0.001,~0.0001$. $t=0$ is time
for the maximum of the waveform modulus and spatial position is the
light ring at $r=3M$ for each value of $\nu$.}\label{fig_converg}
\end{center}
\end{figure}

Furthermore, to validate our result, we check the ``convergence'' of
the $l=m=2$ waveform modulus while $\nu\rightarrow 0$. The result is
displayed in Fig. \ref{fig_converg} and can clearly dedicate that
the waveform we calculated has a good convergence. We think that
this good converging trend, together with the good consistency in
Fig. \ref{fig_Schwz} gives a support to our theoretical idea which
combines EOB dynamics with the Teukolsky perturbation method in the
intermediate mass-ratio inspiral and plunge.

Now, as an application, we show the orbits of the small object
inspiralling into Kerr black holes with different spin:
$a=-0.9,~0.7$. We use the full EOB dynamics equations
(\ref{rdot}-\ref{pfdot}) without the $\nu\rightarrow 0$
approximation. This makes us can contain the conservative mass-ratio
corrections. When $a\neq0,~ \nu\neq0$, it is no possible to write
out an analytical expression for the angular momentum. So we use
Brent's algorithm\cite{Press} to search $L_z$ numerically with a
relative error less than $10^{-13}$. For the Kerr black holes,
because of the so-called frame-dragging effect, the radius of ISCO
is smaller($a>0$) or bigger($a<0$) than the Schwarzschild case. For
example, the extreme Kerr black hole, $a=\pm1$, the ISCO locates at
$r=1M$ and $r=9M$, respectively. And for $a=0.7,~-0.9$,
$r_\text{ISCO}=3.39313,~8.71735$, respectively.

\begin{figure}
\begin{center}
\includegraphics[width=3.0in]{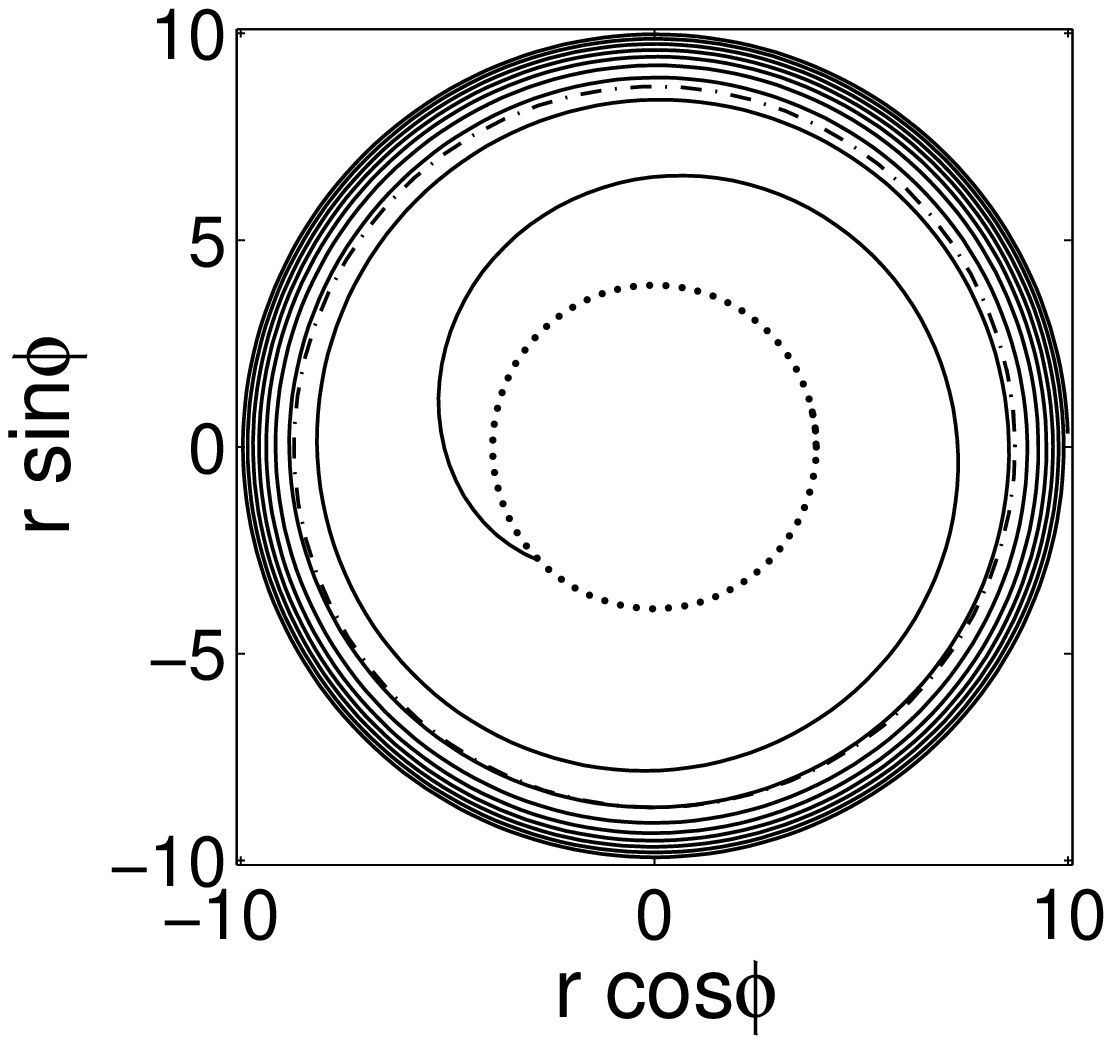}
\includegraphics[width=3.0in]{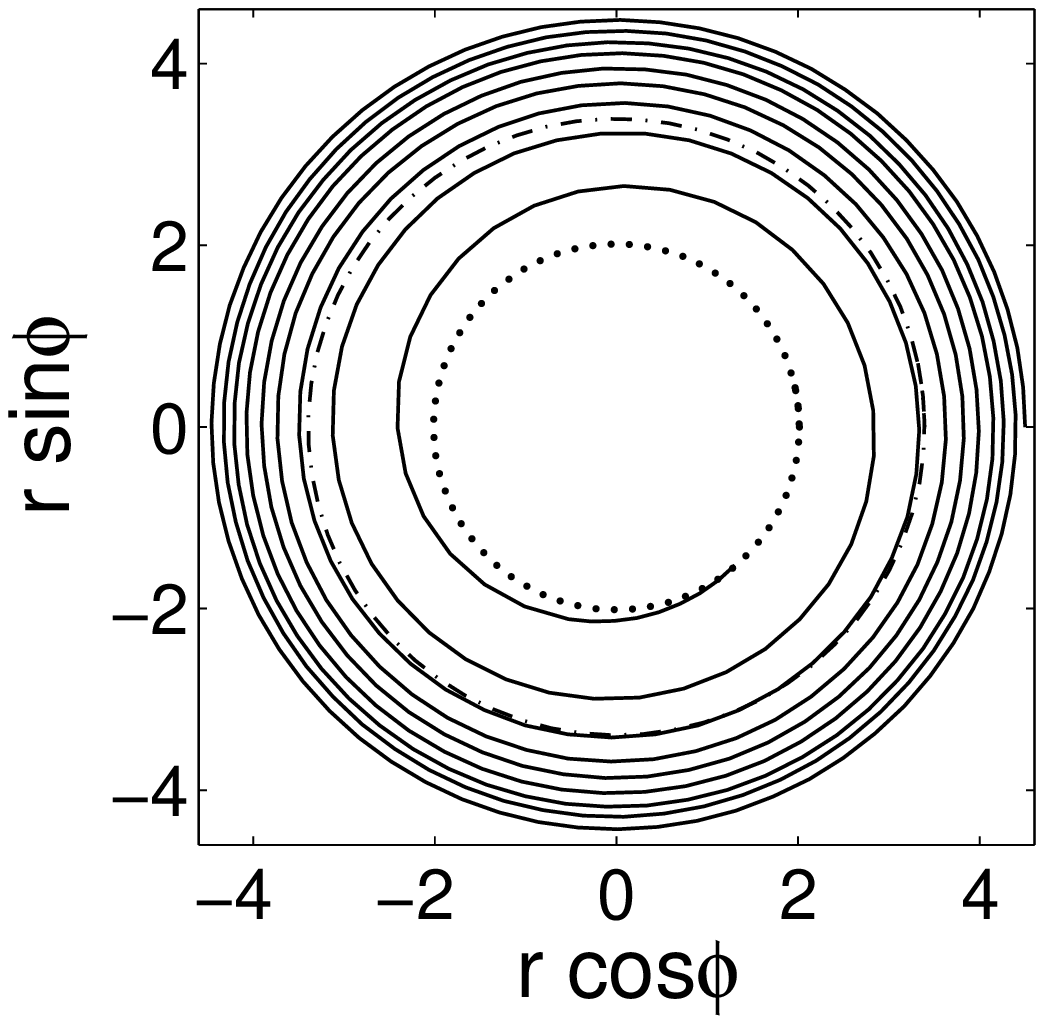}
\caption{Inspiralling and plunge of a small body into a Kerr black
hole with mass ratio $0.01$. The dashed line represents ISCO and
dotted line the light ring. Left panel: retrograde orbit $a=-0.9$
beginning at $r=10M$; right panel: retrograde orbit $a=0.7$
beginning at $r=3.5M$.}\label{fig_inspiral_a}
\end{center}
\end{figure}

For this reason, when $a<0$ (retrograde orbit), the small body
plunges faster into the black hole. At the same time, when $a>0$
(prograde orbit), the small body spirals more cycles before plunge.
These phenomena are clearly displayed in Fig.\ref{fig_inspiral_a}.

\begin{figure}
\begin{center}
\includegraphics[width=3.0in]{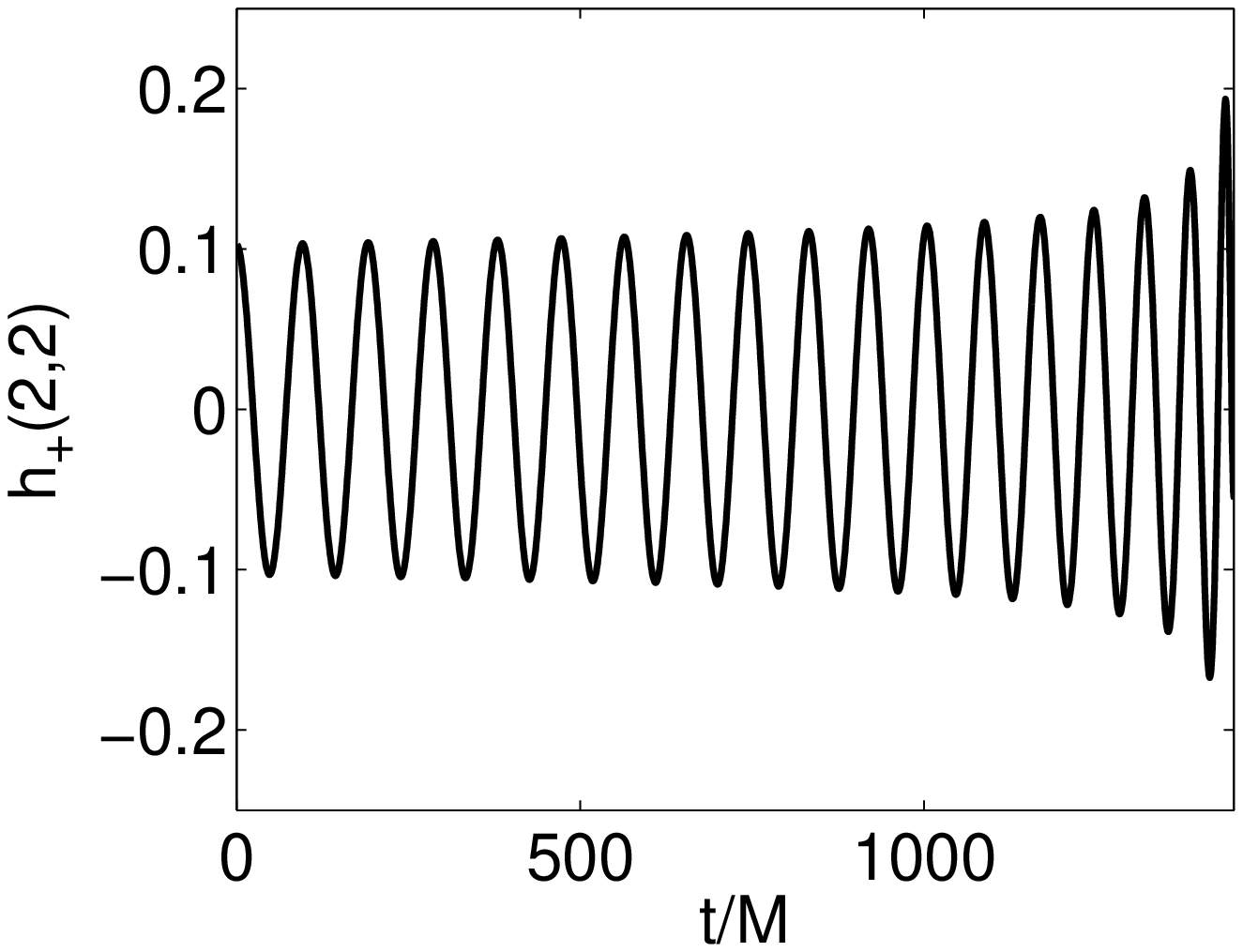}
\includegraphics[width=3.0in]{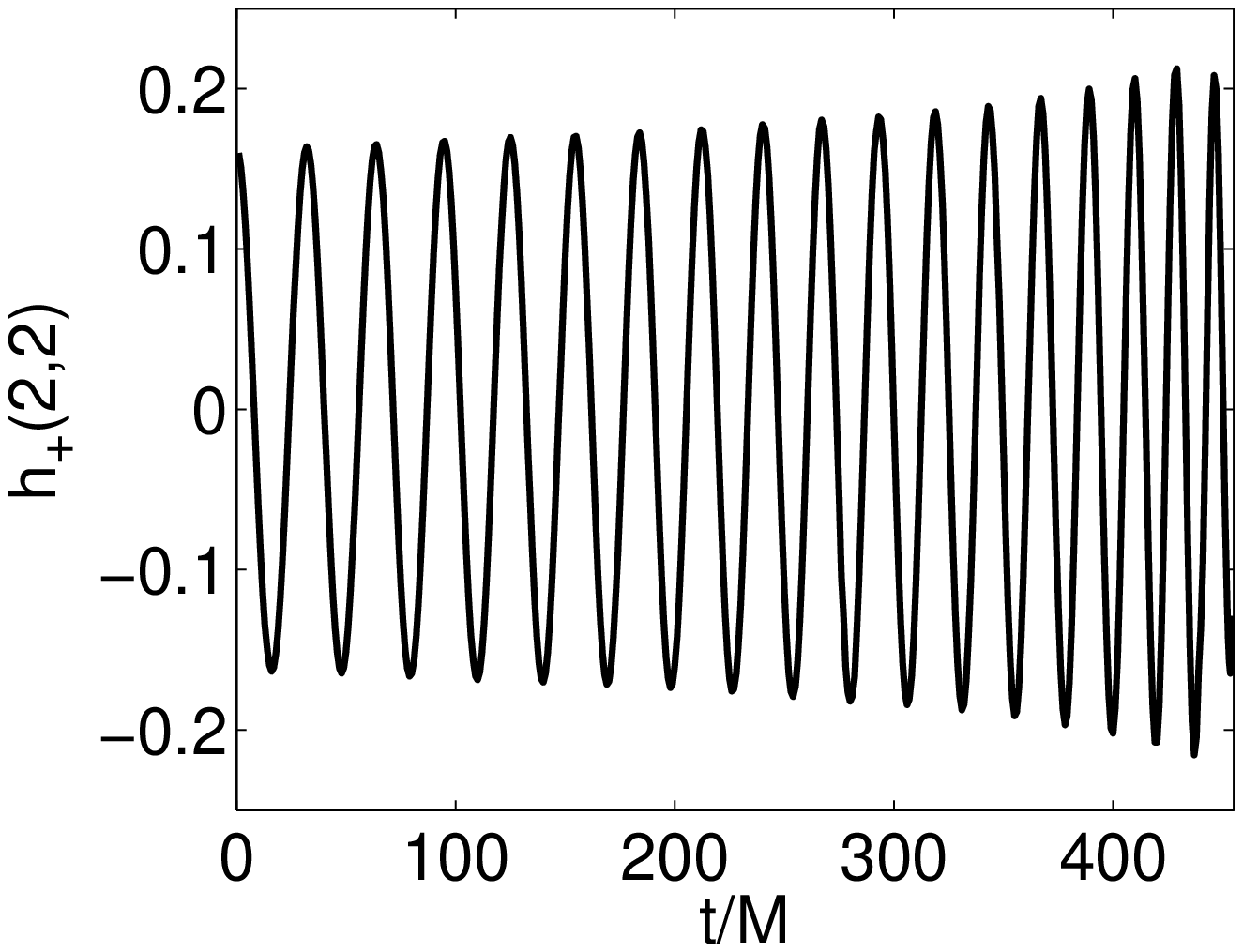}
\caption{GW signal shown through the $l=m=2$ mode of + polarization
corresponding to the dynamics depicted in Fig.\ref{fig_inspiral_a}.
The left panel is the retrograde orbit $a=-0.9$ and the right one
retrograde orbit $a=0.7$.}\label{fig_waveform_a}
\end{center}
\end{figure}

In addition the waveforms are modified by different values of spin
of Kerr black holes. As an example, we show the waveforms of
$a=-0.9,~0.7$ cases in Figure \ref{fig_waveform_a}. After passing
the light ring, based on the analysis in Ref.\cite{Bernuzzi2}, the
merge actually has started and gone into ringdown phase, the energy
flux and waveform cannot be calculated by the frequency-domain
method but can be connected by a quasi-normal mode or time-domain
Teukolsky solution. For this reason, we stop the evolution just
after the small object passing the light ring.

\section{Conclusions and discussion}

Previous works on IMRI dynamics through the EOB method are based on
approximated analytical waveforms. We propose an alternative method
to set the back reaction term for the EOB dynamics via numerically
calculated energy flux directly. We have presented our numerical
results based on the EOB dynamics and the frequency-domain Teukolsky
flux in this paper. This is the first try to use the frequecy-domain
Teukolsky frame in the plunge phase of IMRIs by the help of the EOB
dynamics. Our results are consistent to the previous results using
the analytical energy fluxes of post-Newtonian approximation.

And the results shown here are reasonable. By comparing with the
full EOB evolution in deformed Schwarzschild space-time, a quite
good coincidence before ISCO gives a support for applying the
Teukolsky equation up to mass-ratio $0.01$. After ISCO, the plunge
stage, a small difference comes out due to two different energy
fluxes used.

Since there are few results of post-Newtonian waveforms on elliptic,
and even non-equatorial orbits, it's not easy to generalize the
previous EOB dynamics works to these systems. In principle, our
numerical energy-flux method is not restricted by this. The radial
and polar frequencies can be obtained by evolving an ``imaginary''
geodesic orbit at every step. The ``imaginary'' geodesic orbit is
the solution of Eqs.(19)-(22) but $\hat{\cal F}=0$ in (22). This
solution involves ordinary differential equations with only few
variables, so the computation time is negligible compared with the
total evolution time. Of course, when we try to expand our method to
the elliptic, and even non-equatorial orbits, the numerical cost
will be more high. However, for EMRIs, there were several works
which evolved generic orbits with solving the Teukolsky equation
\cite{Hughes1,Fujita3,Glampedakis1,Glampedakis2}. For IMRIs, we can
afford the numerical cost even for evolving a few tens of orbits.
And more, we can generalize our analysis to generic spinning objects
moving along generic elliptic, and even non-equatorial orbits. These
would be our future works.

The adiabatic approximation works well in inspiral with extreme
mass-ratio limit, but is violated in plunge stage. At the same time,
the EOB dynamics can be used in both inspiral and plunge evolution.
And then the energy fluxes also are calculated by the EOB-PN
expanded waveforms usually. In this traditional method, the EOB
fluxes are based on the post-Newtonian approximation, and losses
accuracy in the high relativistic region. What we do in this paper
is, in the EOB dynamics frame, use the Teukolsky-based energy fluxes
and waveforms to replace the EOB ones. In other words, we adopt the
frequency-domain Teukolsky-based energy fluxes and waveforms but
abandon the adiabatic approximation; we employ the EOB dynamics to
evolve the binary systems but do not use the EOB energy fluxes for
back reaction.

This EOB plus Teukolsky frame we develop in this work can provide an
alternative way to research the gravitational waves from IMRIs. The
dynamical evolution and waveforms for several different spinning
black holes are given in the above sections. But more detailed
analysis about this EOB plus Teukolsky frame should be done farther
in future, when more interesting results can be obtained.

\section*{Acknowledgments}
We thank Dr. R. Fujita and Dr. S. Bernuzzi for valuable discussions.
Z.~Cao was supported by the NSFC (No.~10731080 and No.~11005149).

\end{document}